# Kinetics studies on κ to β-Ga2O3 phase transformations via in-situ high temperature X-ray diffraction

[1]Jingyu Tang, [1]Po-Sen Tseng, [1]Kunyao Jiang, [1]Rachel C. Kurchin, [1]Robert F. Davis and [1]Lisa M. Porter

[1]Department of Materials Science and Engineering, Carnegie Mellon University, Pittsburgh, Pennsylvania 15213, USA

## Abstract

The kinetics of the κ to β-$Ga_2O_3$ phase transformation were investigated in five batches of nominally phase-pure κ-$Ga_2O_3$ thin films heteroepitaxially grown on c-plane sapphire, with film thickness ranging from 700 to 1100 nm, using in-situ high-temperature X-ray diffraction. Phase fractions were quantitatively extracted through modified Rietveld refinement that accounts for preferred orientation, and the transformation kinetics were analyzed using the Johnson-Mehl-Avrami-Kolmogorov (JMAK) model. The applicability of the JMAK model to thin-film materials was evaluated and its lower and upper bounds for thin films and bulk materials were established. Based on this analysis, a method specifically suited for thin-film kinetic studies was developed and yielded reproducible and robust results across all five sample batches. The results indicate that the κ → β phase transformation in ~700-1100 nm films is best described as an interface-controlled, site-saturated nucleation with thickness-limited or effectively two-dimensional growth.

## 1. Introduction

κ-$Ga_2O_3$ is generally considered to be the second most stable $Ga_2O_3$ polymorph after the thermodynamically stable β phase, based on density functional theory (DFT) comparisons of relative free energies among the four commonly accepted polymorphs (α, β, κ, and γ) from 0 K up to 1400 K.[1,2] Experimentally, κ-$Ga_2O_3$ films remain thermally stable up to ~825 °C upon post-deposition annealing,[3–6] whereas the α and γ films of comparable thickness begin to transform into the β phase below 650 °C.[6–9] Higher thermal stability makes κ-$Ga_2O_3$ more attractive for device applications in extreme environments than other metastable polymorphs.

As the only non-centrosymmetric $Ga_2O_3$ polymorph, κ-$Ga_2O_3$ is predicted to exhibit a strong spontaneous polarization along the c-axis in the range of 23–26 μC/cm²,[10–12] which is an order magnitude higher than GaN and AlN. DFT calculations indicate a double-well potential between two opposite polar states, which establishes the thermodynamic possibility of polarization switching.[13,14] However, experimental verification of ferroelectric behavior remains controversial,[14–16] possibly due to the presence of in-plane rotational domains (RDs) within the κ films. These RDs lead to topologically constrained polarization domain well propagation, thereby suppressing or obscuring macroscopic polarization reversal.[17] In contrast, polarization-induced two-dimensional electron gas (2DEG) formation does not require switchable polarization but instead relies on the presence of spontaneous polarization and a polarization discontinuity at heterointerfaces. This enables the formation of high-density, dopant-free, 2DEGs in κ-$(Al_xGa_{1-x})_2O_3/Ga_2O_3$-based high-electron-mobility transistors, with carrier densities estimated up to ~$10^{14}$ cm$^{-2}$,[13] an order of magnitude higher than modulation doped β-($Al_xGa_{1-x}$)$_2$O$_3$/Ga$_2$O$_3$ heterostructures.[18–20] Nevertheless, experimental demonstrations remain limited, possibly due to poorly controlled interfaces. The coexistence of κ and β phases at the interface has been observed,[21–26] forming a rough and non-abrupt interface that significantly degrades electron density and mobility through grain boundary-related scattering.

While the strained-induced stabilization effect on the coexistence of κ and β phases at interfaces and their evolution during film growth have been investigated,[23,24,26] the kinetics of such phase transformations including transition rates, activation energies, and the underlying nucleation and growth mechanisms of β phase within κ phase are not fully understood. In this work, we investigate the κ to β phase transformation kinetics in epitaxial films using in-situ high-temperature X-ray diffraction (HT-XRD). The Johnson-Mehl-Avrami-Kolmogorov (JMAK) model is applied to examine the transformation kinetics during isothermal annealing. The Avrami exponents and effective rate constants are extracted at different temperatures, and the corresponding activation energy is determined. The applicability of the JMAK model to finite-thickness thin-film systems is also discussed.

## 2. Methods

Five batches of nominally phase-pure κ(ε)-Ga$_2$O$_3$ films, with thicknesses ranging from 690 nm to 1100 nm, were heteroepitaxially grown on chemo-mechanically polished, epi-ready (0001) sapphire (Al$_2$O$_3$) substrates (with a 0.15° offcut towards the m-axis) at 530 °C using a vertical, low-pressure, cold-wall metal-organic chemical vapor deposition reactor at a growth rate of ~760 nm/h. Triethylgallium (TEGa) and ultra-high-purity O$_2$ (5N) were used as the Ga precursor and oxidant, respectively. The molar flow rate of the TEGa were in the range of 13.3 to 70.2 μmol/min, whereas the O$_2$ molar flow rate were fixed at 2.23×10$^{-2}$ mol/min and the chamber pressure was maintained at 20 torr. These growth conditions were optimized previously and further details on process tuning and characterization of similar films have been reported in our previous publications.[16,27,28] Three films of ~720 nm were deposited in the same deposition run, while the other two films with thicknesses of ~690 nm and ~1100 nm were sequentially grown under the same growth conditions. A β-Ga$_2$O$_3$ film was also grown at 830 °C and used as a reference sample for the β phase.

High-resolution X-ray diffraction (HR-XRD) patterns of the as-grown and annealed films were acquired using a Panalytical X'pert Pro MPD X-ray diffractometer equipped with a 4-crystal Ge (220) monochromator. Isothermal HT-XRD patterns were recorded using a Malvern Panalytical Empyrean diffractometer equipped with an HTK 16N high-temperature chamber. The latter instrument was employed to investigate the κ to β phase transformation at five fixed annealing temperatures: 810 °C, 820 °C, 830 °C, 840 °C and 850 °C, respectively, in air until κ phase was undetectable and no further increase of the β phase diffraction intensity was observed.

The transformed volumetric fraction $f_v(t)$ of the new phase (β phase) was quantified across five batches of sample using a modified Rietveld refinement approach. Because both the parent κ phase and the product β phase exhibit strong preferred orientation, conventional Rietveld refinement alone is insufficient for quantitative phase analysis and volumetric fraction extraction. To account for texture effects, the March-Dollase (MD) function was incorporated into the refinement procedure. The preferred orientations were identified as (004) for the κ phase and ($\bar{2}01$) for the β phase. At each isothermal annealing temperature, $f_v(t)$ was extracted using the GSAS-II software package by minimizing the weighted-profile R-factor ($R_{wp}$) between the experimentally observed and calculated diffraction patterns. A detailed description of the data extraction procedure based on the modified Rietveld refinement approach is provided in Supplementary Material Section I. The transformation kinetics were analyzed using the JMAK formalism as a parametric description of the measured $f_v(t)$, from which the effective rate constant $k$ and Avrami exponent $n$ were extracted at each temperature. The activation energy $E_a$ for the phase transformation was subsequently obtained from an Arrhenius plot of $\ln k$ versus inverse temperature ($1/T$).

## 3. Results and Discussion

### 3.1 Orientation relationships before and after the phase transformation

Figure 1(a) shows a contour map of the in-situ HT-XRD patterns recorded during a complete κ to β phase transformation performed at 830 °C on a 720 nm thick film. The x-axis corresponds to 2θ angles ranging from 15° to 45°, the y-axis represents the annealing time, and the color scale represents the diffracted X-ray intensity. The (0006) sapphire peak was used as an internal reference for scan alignment. No intermediate phase was detected throughout the transformation, as the disappearance of the (002) κ-$Ga_2O_3$ family of planes is directly accompanied by the emergence of the ($\bar{2}01$) β-$Ga_2O_3$ family of planes, with no additional diffraction peaks observed. These in-situ HT-XRD results are in good agreement with in-situ transmission electron microscopy (TEM) observations reported by Cora *et al*,[3] who observed a gradual structural reorganization of the κ phase initiating at ~650 °C through domain-size coarsening, followed by a rapid exothermic transformation to β at ~950 °C without any detectable intermediate phase.

Figure 1(b) shows the corresponding ex-situ HR-XRD 2θ-ω scans of the same film before and after in-situ HT-XRD annealing at 830 °C. After annealing, the transformed β (from κ) film exhibits the same diffracted peak positions as an as-deposited β film directly grown on c-plane sapphire at 830 °C. This indicates that the transformed β phase maintains the same out-of-plane epitaxial relationship with the substrate as an as-deposited β film. Such epitaxial consistency suggests a close structural similarity of the oxygen sublattice among the (0006) sapphire, ($\bar{2}01$) β[29–31] and (002) κ planes,[27,32,33] which will be discussed in Fig. 2.

Figure 1(c) compares both ($\bar{2}01$) symmetric and ($\bar{4}01$) non-symmetric X-ray rocking curves (XRCs) of the transformed β film and the as-deposited β film. The post-deposition annealed β film exhibits a narrower full-width at half maximum (FWHM) than the as-deposited β film, indicating the reduction of the dislocation density. This can be attributed to β phase inheriting the epitaxial registry of the κ phase, which already grew epitaxially on the substrate with an FWHM approximately one-half that of the as-grown β film, reducing mosaicity. As shown in Figs. 1(a)-(b), the κ to β phase transformation proceeds with a specific orientation relationship, which provides a guided pathway for the transformed β phase to align more effectively than the randomly nucleated β film in the as-deposited film.

The in-plane epitaxial relationships before and after transformation are shown in the XRD φ-scans in Fig. 1(d). Three in-plane RDs of κ before annealing and six in-plane RDs of β after transformation are observed. According to group theory,[34] the number of RDs can be predicted from the interfacial symmetrical mismatch using

$$N_{\text{RD}} = \frac{lcm(n,m)}{m}, \quad (1)$$

where $N_{\text{RD}}$ is the expected number of RDs in the epilayer, $n$ is the rotational symmetry of the substrate surface, and $m$ is the rotational symmetry of the epilayer, both defined with respect to the surface normal. The term $lcm$ refers to the least common multiple. For c-plane sapphire, $n = 6$ (sixfold symmetry along [0001]). For κ-Ga₂O₃, $m_\kappa = 2$ (twofold symmetry along [001]), and for β-Ga₂O₃, $m_\beta = 1$ no rotational symmetry along [102]). Therefore, $N_{\text{RD}} = 3$ for κ on c-plane sapphire and $N_{\text{RD}} = 6$ for β on c-plane sapphire, in agreement with experimental observations. The φ-scan of the transformed β phase exhibits identical peak positions to those of the as-deposited β film, indicating that the transformed β is both out-of-plane and in-plane aligned with the as-deposited β directly grown on c-plane sapphire.

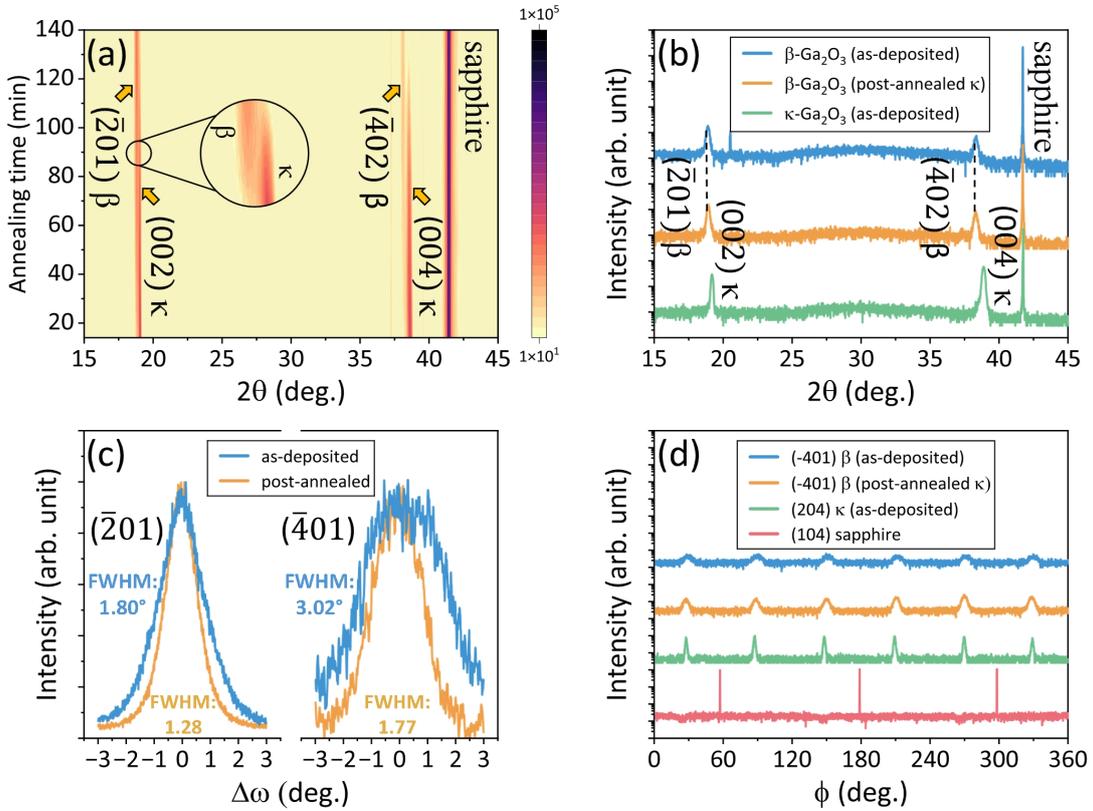

Figure 1. (a) In-situ HT-XRD contour map showing the κ to β-Ga₂O₃ transformation performed at 830 °C in air. (b) Ex-situ HR-XRD 2θ-ω scan of the same κ-Ga₂O₃ film before

and after annealing at 830 °C. (c) Comparison of the ($\bar{2}$01) symmetric and ($\bar{4}$01) non-symmetric XRCs between the transformed β film and the as-deposited β film. (d) XRD φ-scans of the κ-$Ga_2O_3$ film prior to annealing and the resulting β-$Ga_2O_3$ after annealing.

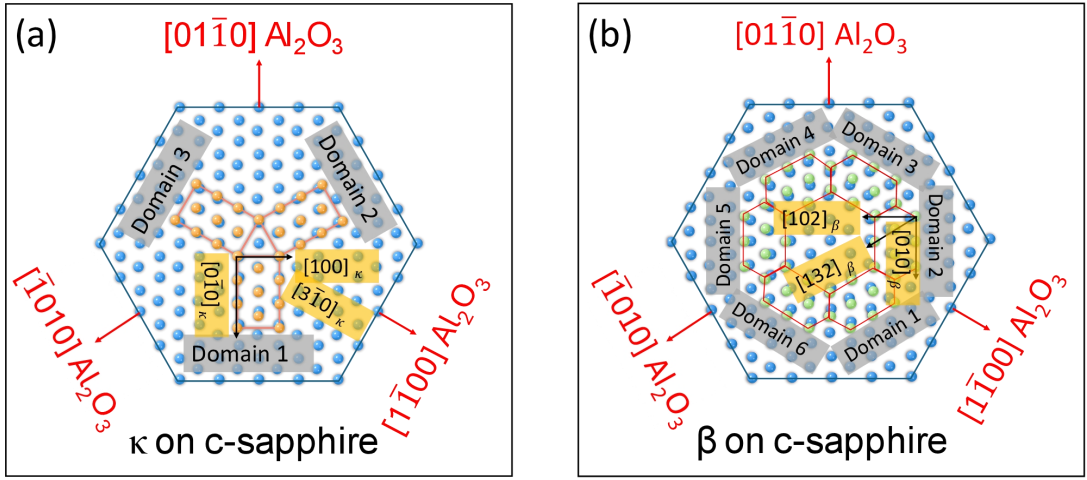

Figure 2. Schematic illustrations of the in-plane oxygen atomic arrangement between the film and substrate and the presence of rotational domains for (a) the as-deposited κ film and (b) the β film transformed from κ after annealing.

The presence of in-plane RDs in both as-grown κ and β transformed from κ, along with their orientational relationships with c-plane sapphire, are illustrated in Fig. 2. The summarized in-plane orientational relationships are as follows:

For κ-$Ga_2O_3$:

$[3\bar{1}0]_\kappa$ (domain 1) // $[\bar{3}\bar{1}0]_\kappa$ (domain 2) // $[010]_\kappa$ (domain 3) // $[1\bar{1}00]_{sapphire}$

For β-$Ga_2O_3$ (transformed from κ):

$[\bar{1}\bar{3}\bar{2}]_\beta$ (domain 1) // $[\bar{1}3\bar{2}]_\beta$ (domain 2) // $[010]_\beta$ (domain 3) // $[132]_\beta$ (domain 4)

// $[1\bar{3}2]_\beta$ (domain 5) // $[0\bar{1}0]_\beta$ (domain 6) // $[1\bar{1}00]_{sapphire}$

## 3.2 Finite thickness and growth anisotropy effects on JMAK model

To quantitively analyze the transformation kinetics, we have applied the JMAK model:

$$f_v(t) = 1 - \exp(-(kt)^n), \qquad (2)$$

where $f_v(t)$ is the volume fraction of the transformed phase as a function of time (t), $k$ is the effective (temperature-dependent) transformation rate constant, and $n$ is the Avrami exponent associated with the nucleation and growth mechanism of the transformed phase.[35] In this study the κ to β phase transformation occurs in a plate-like (thin-film) geometry, where the growth of nuclei of the transformed phase is spatially confined along the film normal direction. Once the growth front reaches either the top surface (film-free surface) or the bottom surface (film-substrate interface), continued growth can proceed only through lateral expansion within the film plane. Such geometric confinement violates one of the core assumptions of the classical JMAK model, namely the infinite and unrestricted growth of the nuclei in all directions.

To account for the finite film thickness and its effect on transformation kinetics, it is necessary to revisit the original derivation of the JMAK model and explicitly examine the thickness-dependent effects on the evolution during the phase transformation. The JMAK model shown in Eq. (1) is in fact a specific solution derived from the original Kolmogorov formulation:[36,37]

$$X(t) = 1 - \exp\left[-\Phi(t)\right], \qquad (3)$$

Where $X(t)$ is the actual volumetric fraction that has transformed, accounting for impingement effects, and $\Phi(t)$ is the extended volume fraction that would be transformed if all nuclei grew freely without impingement between growing domains. $\Phi(t)$ is determined by integrating the nucleation rate (or nucleation density) and the growth of individual nuclei over time, which explicitly depends on the dimensionality of the growth and the geometry of the system. The exponential relationship arises from the probabilistic treatment of randomly distributed nucleation events. The form $\Phi(t) = (kt)^n$ is only valid when the system is effectively infinite compared with the size of an individual grain.

For a thin film, $\Phi(t)$ must be modified to account for the finite thickness effect (see the discussion in Supplementary Material Section II). Here we examine the combined effects of film thickness $h$ and growth anisotropy $(v_\text{in}/v_\text{out})$ on the effective Avrami exponent $n$, where $v_\text{in}$ and $v_\text{out}$ denote the lateral (in the plane of the film) and vertical (along normal to film surface) growth velocities, respectively. Several assumptions are adopted to simplify the analysis. For continuous nucleation, where new nuclei continue to form throughout the transformation process, the nucleation rate $\dot{N}_v$ is assumed to be constant. For site-saturated nucleation, where all nuclei are formed and rapidly saturated at the onset of transformation and no new nuclei form afterward, the density of nuclei $N_v$ is assumed to be constant. In both cases, the growth velocities of the transformation front along the x-, y-, and z-directions (i.e., $v_x$, $v_y$, and $v_z$) are taken to be constant, implying interface-controlled growth rather than diffusion-controlled kinetics. Accordingly, the phase transformation proceeds mainly through local atomic rearrangement or shear-like reconstruction without long-range atomic diffusion,[3,6] and the rate-limiting step is governed by the kinetics of the moving interface between the parent and product phases.

Under these assumptions, the dependence of the effective Avrami exponent $n$ on film thickness and growth anisotropy is illustrated in Figs. 3(a-c). A detailed mathematical derivation is provided in Supplementary Material Section II. The x-axis represents the dimensionless thickness, defined as $h/\lambda$, where $\lambda$ is the characteristic length of the transformation. For continuous nucleation, $\lambda = \left(\dot{N}_v/v\right)^{-1/4}$, where $v = (v_x v_y v_z)^{1/3}$ is the geometric mean of the growth velocity. For site-saturated nucleation, $\lambda = N_v^{-1/3}$. The y-axis represents the effective Avrami exponent $n$. Three anisotropy ratios ($v_{in}/v_{out} = 10, 1, 0.1$) are considered. All three cases (Figs. 3(a-c)) show the same general trend. When the film is sufficiently thin, individual nuclei rapidly impinge on the top and bottom film boundaries, after which further growth is confined to the lateral directions, corresponding to two-dimensional (2D) growth. In this limit, the effective Avrami exponent approaches 3 for continuous nucleation and 2 for site-saturated nucleation and becomes nearly independent of film thickness. For intermediate thicknesses, nuclei initially grow in three dimensions (3D growth) until reaching the film boundaries, followed by lateral 2D growth. As a result, the effective Avrami exponent varies continuously with film thickness, reflecting the crossover from 3D to 2D growth. In this regime, a global or effective Avrami exponent only represents an average kinetics over the entire transformation interval. Instead, a local or instantaneous Avrami exponent should be evaluated to capture different kinetic stages of the transformation. As the film thickness further increases, the film approaches the bulk limit. In this regime, nuclei grow freely in 3D over nearly the entire transformation interval, yielding a constant Avrami exponent of 4 for continuous nucleation and 3 for site-saturated nucleation and becomes nearly independent of film thickness. These values are consistent with the classical JMAK predictions for bulk materials.[36,37]

Figure 3(b) shows the isotropic growth case ($v_{in}/v_{out}$ = 1). The lower and upper bounds of the 2D- and 3D- dominated regimes are $h/\lambda \cong 0.2$ and 5, respectively. These values are comparable to those reported from level-set computational simulations performed independently of the JMAK model.[35,38] The agreement supports the validity of the analytical solutions derived in this work based on the original Kolmogorov formulation and the appropriate modification on the $\Phi(t)$ term. The effect of growth anisotropy can be observed by comparing Figs. 3(a-c). When lateral growth is faster (Fig. 3(a)), both curves shift leftward relative to the isotropic case. This shift indicates that, for a fixed film thickness and by the time individual nuclei impinge on the film boundaries, a larger fraction of the film volume has transformed, leaving less parent phase to be transformed via lateral 2D growth, thus the overall transformation kinetics more closely resemble 3D growth throughout the transformation interval. Conversely, when the lateral growth is slower (Fig. 3(c)), a smaller fraction of the volume has transformed, leaving larger parent phase to be transformed through lateral 2D growth, leading to a rightward shift of the curves relative to the isotropic case.

Experimentally, determining $\lambda$ and thus $h/\lambda$ as well as the degree of growth anisotropy is challenging, since it requires in-situ measurements of the nucleation rate (or nucleation density) and the growth velocities along different directions. Consequently, although these plots show the regimes in which the classical JMAK framework is applicable or not, directly extracting $n$ is nontrivial. Alternatively, evaluating the instantaneous $n$ and tracking its evolution throughout the transformation interval provides a more physically interpretable and experimentally accessible approach. Figures 3(d-f) and 3(g-i) show the instantaneous $n$ as a function of transformed fraction $X(t)$ for the continuous nucleation and site-saturated nucleation cases, respectively, corresponding to the three anisotropy ratios shown in Figs. 3(a-c). The isotropic case shown in Figs. 3(e) and 3(h) are used to illustrate the general trends of these plots. For films with $h/\lambda < 0.2$, the instantaneous $n$ remains nearly constant and approaches the thin film value throughout the entire transformation interval. For $h/\lambda > 5$, $n$ approaches the bulk value predicted by the classical JMAK model. Within the intermediate regime, $n$ gradually converges toward 2D values as the transformation proceeds. Specifically, $n$ approaches 3 in the continuous nucleation and approaches 2 in the site-saturated nucleation. These limiting values correspond to the effective 2D growth regime that dominates once nuclei impinge on the film boundaries and further growth becomes laterally confined. The deviation occurred during the early stage of transformation, when 3D growth still contributes significantly. As the transformed fraction increases and boundary effects become dominant, the instantaneous exponent collapses toward its thickness-limited value. The effect of growth anisotropy can be observed by comparing Figs 3(d-f) and 3(g-i). When lateral growth is faster, the deviation of the instantaneous exponent persists to larger transformed fractions, and the collapse toward the 2D-limited values occurs later. In contrast, when lateral growth is slower, the instantaneous exponent rapidly decreases toward its thickness-limited value at smaller transformed fractions, reflecting the earlier onset of boundary-controlled growth.

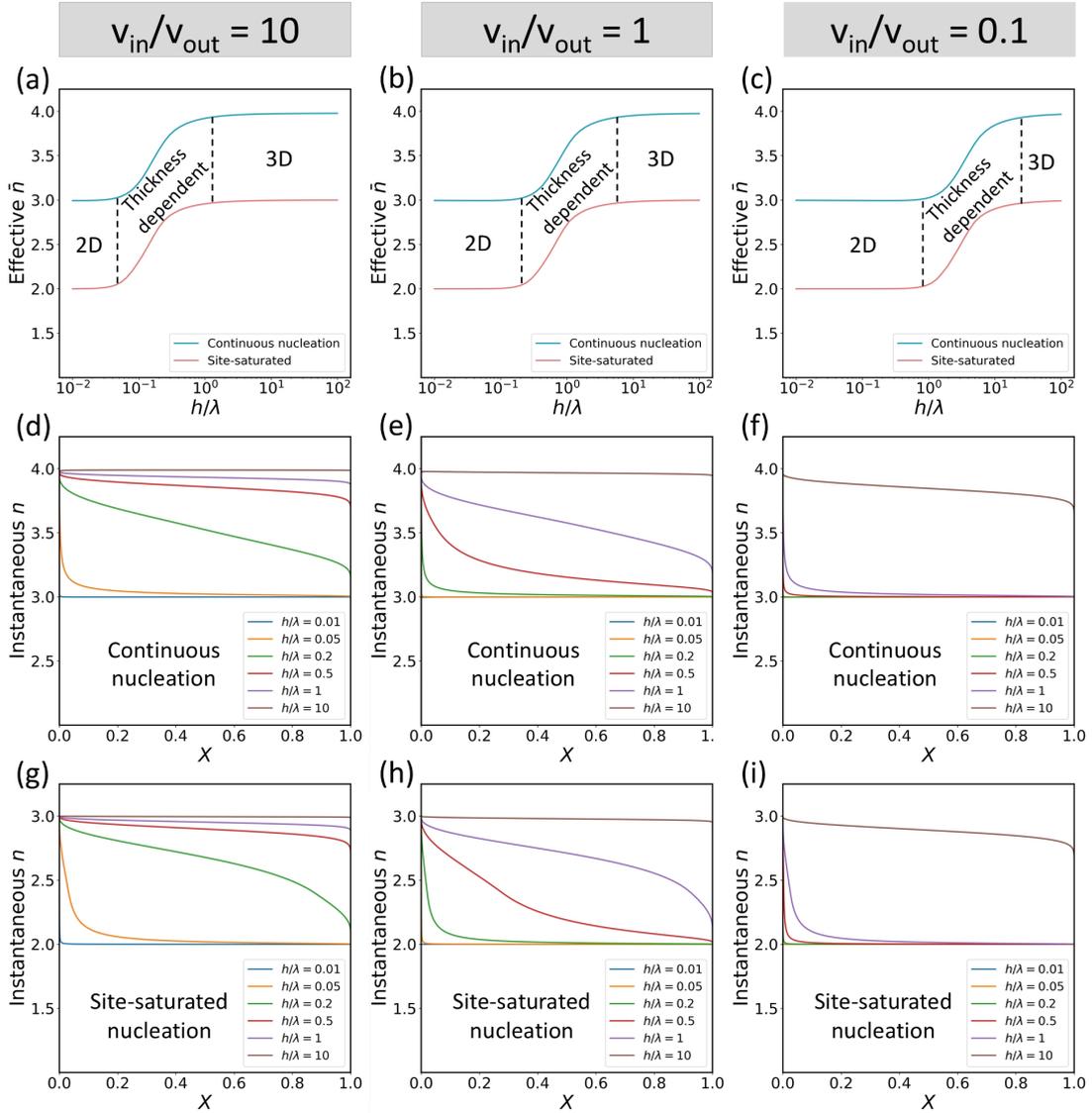

Figure 3. Effective Avrami exponent as a function of the dimensionless thickness $h/\lambda$ for $v_{\text{in}}/v_{\text{out}} = 10$ (a), 1 (b), and 0.1 (c). Panels (d)-(f) show the corresponding instantaneous Avrami exponent versus the volumetric transformed fraction $X(t)$ for the continuous nucleation case. Panels (g)-(i) present the corresponding instantaneous Avrami exponent versus $X(t)$ for the site-saturated nucleation case.

## 3.3 In-situ Isothermal XRD annealing and kinetics investigation

All five κ films deposited on 10 mm × 10 mm c-plane sapphire substrates were each further diced into six identical specimens (3.3 mm × 5 mm) for conducting the in-situ HT-XRD sequential annealing. Owing to the small sizes and the thermal-contact variations on the hot stage, the actual sample temperature may deviate from the set point temperature. Therefore, each specimen was sequentially subjected to in-situ HT-XRD isothermal annealing at 810 °C, 820 °C, 830 °C, 840 °C, and 850 °C, and the sample temperature was calibrated for each measurement using the (0006) sapphire peak position. The detailed temperature calibration procedure can be found in Supplementary Material Section III. The collected datasets were grouped and labeled as Batches 1-5, respectively. Batches 1-3 correspond to three 720 nm films grown during the same growth run, while Batches 4 and 5 correspond to additional 690 nm and 1100 nm films grown under the same growth conditions as Batches 1-3.

Figure 4 shows the HT-XRD contour maps collected at each temperature for Batch 5 (1100 nm). (The corresponding results for the other two batches are provided in Supplementary Material Section IV.) As the annealing temperature increases, the phase transformation rate is accelerated, leading to a shorter time required to complete the transformation. Across the entire temperature range investigated (810 °C-850 °C), a direct κ to β transformation is observed with no intermediate phase detected. Thus, the constraint of $f_\kappa(t) + f_\beta(t) = 1$, where $f_\kappa(t)$ and $f_\beta(t)$ are the volume fractions of κ and β, was imposed during the modified Rietveld refinement process.

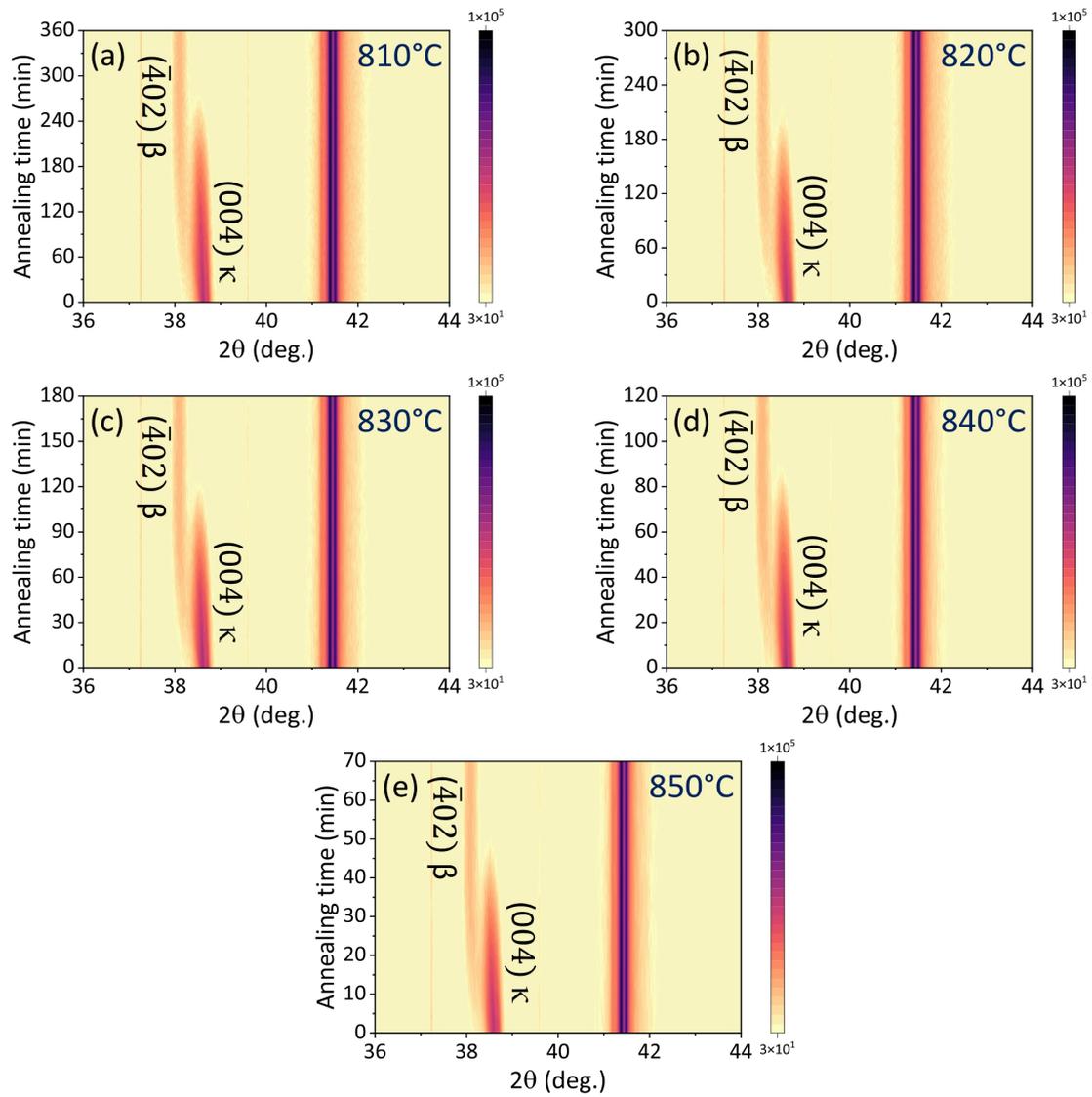

Figure 4. HT-XRD contour maps acquired during in-situ isothermal annealing of κ films (Batch 1) at 810 °C, 820 °C, 830 °C, 840 °C, and 850 °C. The x-axis corresponds to 2θ angles ranging from 36° to 44°, the y-axis represents the annealing time, and the color scale denotes the diffracted X-ray intensity.

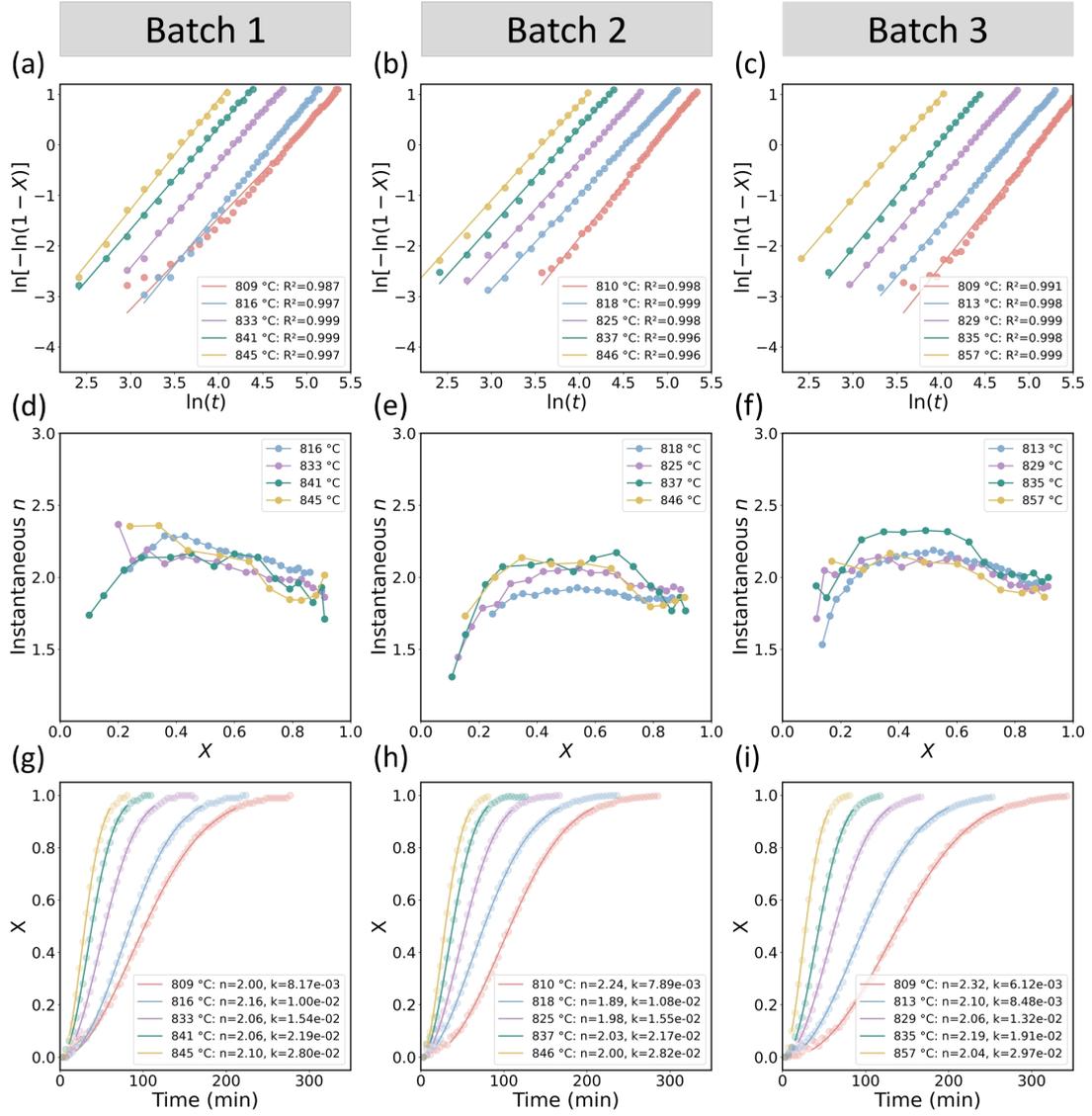

Figure 5. (a–c) Avrami plots for κ-films (Batches 1–3) annealed at different temperatures. Solid lines represent linear fits to the JMAK model. (d-f) Instantaneous Avrami exponent $n$ as a function of transformed fraction $X$. (g-i) Experimentally extracted volumetric transformation fraction $X(t)$ together with the JMAK model fits using the extracted kinetic parameters ($n$ and $k$), demonstrating the overall agreement between the model and experimental data.

Figure 5 shows the κ to β phase transformation kinetics of the three batches analyzed using Avrami plots and JMAK fitting. The Avrami plots in Figs. 5(a-c) exhibit excellent linearity over nearly the entire transformation interval, with correlation coefficients ($R^2$) approaching 1, indicating that the JMAK model with a single $n$ provides a reasonable description of the overall kinetics under each isothermal condition between 810 °C and 850 °C. A slight deviation is observed at ~810 °C during the early stage of the transformation. At this temperature, the transformation rate is the lowest, and the κ-phase peak intensity in the in-situ XRD spectra shows noticeable fluctuations before the monotonic decrease in the initial stage. These fluctuations introduce uncertainty in the determination of the volume fraction, leading to slight deviations from ideal linear behavior in the Avrami plot. Notably, such fluctuations are mainly observed at ~810 °C and are almost absent at higher temperatures, suggesting that this deviation arises from experimental uncertainty rather than a thickness-dependent effect on the extracted Avrami exponent $n$ shown in Figs. 3(a-c). In contrast, at higher temperatures, the transformation proceeds more rapidly, resulting in a monotonic decrease of the κ-phase peak and a monotonic increase of the β-phase peak. The stronger and more dominant peak evolution relative to instrumental noise minimized intensity fluctuations, yielding more reliable volumetric phase fraction extraction and improved linearity in the Avrami plots. The slopes extracted from these linear fits yield Avrami exponents $n$ close to 2 for most temperatures and across all three batches, while the intercepts yield the temperature-dependent transformation rate constant $k$, which systematically increases with annealing temperature, consistent with a thermally activated interface-controlled transformation.

Figures 5(d-f) show the instantaneous $n$ versus the volumetric transformed fraction $X(t)$ for Batches 1-3. Although $n$ is not strictly constant over the entire transformation process, it remains tightly clustered around 2 (within ~ ±10%) and appears nearly flat throughout the transformation interval of $X(t) = 0.1$-$0.9$. This behavior indicates that the dominant transformation mechanism remains essentially unchanged. Such a trend is consistent with the site-saturated nucleation case where the thickness is below or near its 2D limit shown in Figs. 3(g-i).

The $X(t)$ verses $t$ curves in Figs. 5(g-i) compare the experimentally extracted transformation fraction versus time, and the corresponding fits using the JMAK model with parameters extracted from Figs. 5(a-c). The fitted curves, obtained using a single Avrami exponent $n$ (with $n \approx 2$ for all batches) together with the temperature-dependent rate constant $k$, closely reproduce the sigmoidal kinetics observed experimentally. Combined with the results shown in Figs. 5(d-f), it can be concluded that the extracted Avrami exponent $n \approx 2$ is not only an effective, weighted-average fitting parameter, but also reflects the instantaneous exponent, which remains nearly constant and independent of the transformed fraction $X(t)$ throughout most of the transformation process. These findings indicate that the κ to β phase transformation in films with a thickness of ~700 nm is dominated by a 2D confined phase transformation mechanism with site-saturated nucleation, where β nuclei are uniformly distributed within the κ film and form at the onset of the transformation (or are already present in the as-grown film). The possibility of surface-dominated nucleation can be ruled out.

For homogeneous site-saturated nucleation in 2D-confined films, as represented in Figs. 3(g-i), the instantaneous $n$ rapidly decreases from 3 to 2 and then remains at 2 over nearly the entire transformation interval, consistent with the experimental observation shown in Figs. 5(d-f). In contrast, level-set simulations predict that, for surface nucleation, $n$ initially rises above 3 and only decreases toward 2 after nuclei growing from one surface impinge on the opposite surface.[38] We do not observe such behavior.

In-situ TEM studies performed by Cora *et al.*[3] further support our findings, showing that the transformation initiates within the interior of the κ film and no additional nuclei form during the subsequent phase transformation. The self-consistent and reproducible results of such behavior across Batches 1-3 demonstrate the robustness of the kinetics analysis. Two additional samples with film thicknesses of ~720 nm and ~1100 nm were subjected to the same kinetic analysis and yielded similar results compared to those from Batches 1-3. The results on these two samples are provided in Supplementary Material Section V.

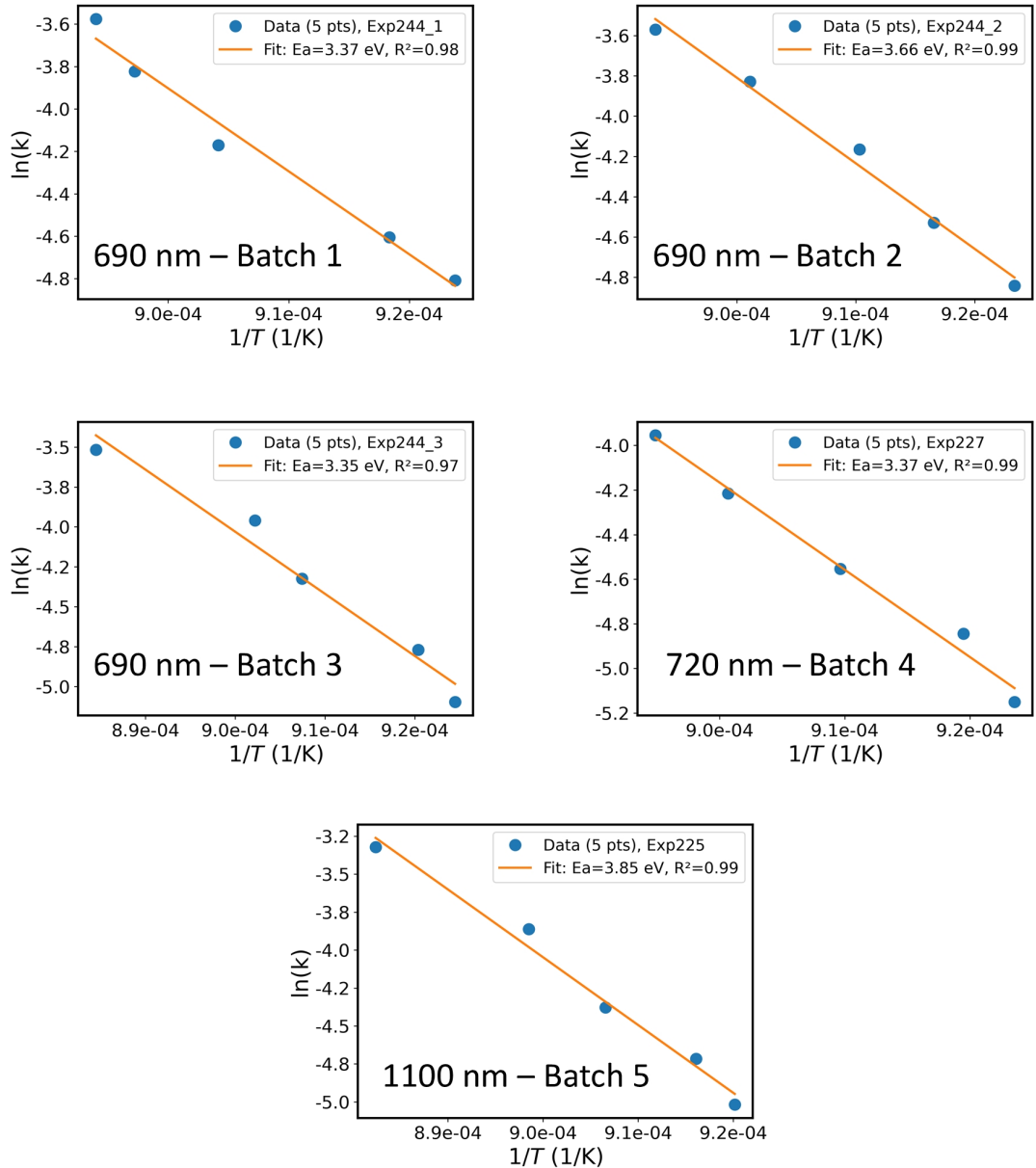

Figure 6. Arrhenius plots of the temperature-dependent rate constant $k$ extracted from the JMAK fitting for Batches 1-3 (~730 nm), together with Batch 4 (690 nm) and Batch 5 (1100 nm) films.

Based on the temperature-dependent rate constants $k$ extracted from the JMAK fitting, the activation energy ($E_a$) for the κ to β phase transformation was determined using the Arrhenius plot, as shown in Fig. 6 for Batches 1-3 and for the 720 nm and 1100 nm films. Linear fitting of $\ln(k)$ versus $1/T$ yields activation energies of 3.37 eV, 3.66 eV, and 3.35 eV, 3.37 eV and 3.85 eV for Batches 1, 2, and 3 and 720 nm and 1100 nm films, respectively. The linearity confirms that the transformation kinetics follow Arrhenius behavior over the investigated temperature range. The extracted activation energies are in close agreement across the three independent batches and between the films with different thicknesses, further demonstrating the reproducibility and internal self-consistency of the kinetic analysis. The small variation in $E_a$ among batches likely reflects minor experimental uncertainties rather than intrinsic differences in the transformation mechanism.

## 4. Conclusion

In summary, the κ to β phase transformation in five batches of nominally phase-pure κ-$Ga_2O_3$ thin films has been systematically investigated through in-situ isothermal HT-XRD. The transformation proceeds directly from κ to β without the formation of any detectable intermediate phase, while preserving both out-of-plane and in-plane epitaxial relationships with the c-plane sapphire substrate consistent with an as-grown β film.

By revisiting the original Kolmogorov formulation and explicitly accounting for finite thickness and growth anisotropy, we clarify the applicability of the classical JMAK model in these films. The effective Avrami exponent depends on the dimensionless thickness $h/\lambda$ as well as the growth anisotropy ratio. Experimentally, Avrami plots for all 5 batches show excellent linearity over the transformation interval, yielding a consistent Avrami exponent of $n \approx 2$ across temperatures from 810 °C to 850 °C. The instantaneous $n$ analysis further confirms that the exponent remains nearly constant throughout the transformation interval of 0.1-0.9 in volume fraction, indicating that $n \approx 2$ is not only a fitting average but reflects the intrinsic kinetic behavior. The extracted rate constants exhibit clear Arrhenius behavior with activation energies of 3.35-3.85 eV. The strong agreement among the five batches demonstrates the robustness and reproducibility of the analysis. Overall, the κ → β transformation in ~700-1000 nm films can be best described as interface-controlled, site-saturated nucleation with thickness-limited or effectively 2D growth.

## Supplementary Material

See the supplementary material for the detailed mathematical derivation of the modified JMAK model and additional data.

## Acknowledgments

This material is based upon work supported by the Air Force Office of Scientific Research (Program Manager, Dr. Ali Sayir) under Award No. FA9550-21-1-0360, by the II-VI Foundation, and by the National Science Foundation (Program Manager: Yaroslav Koshka, Grant No. DMR-2324375).The use of the Materials Characterization Facility at Carnegie Mellon University was supported by Grant No. MCF-677785.

## Conflict of Interest

The authors have no conflicts to disclose.

## Data Availability

The data that support the findings of this study are available from the corresponding author upon reasonable request.